\documentclass[twocolumn,showpacs,preprintnumbers,amsmath,amssymb,a4paper,superscriptaddress]{revtex4}
\usepackage{amsmath}
\usepackage{graphicx}
\usepackage{epstopdf} 
\usepackage{dcolumn}
\usepackage{bm}

\begin{document}
\title{A single electron transistor with quantum rings}
\date{\today}
\author{Ali Hosseinzadeh}
\email[]{hosseinzadeh@iasbs.ac.ir;a.aghdam67@gmail.com}
\author{Shahpoor Saeidian}
\email[]{saeidian@iasbs.ac.ir}
\affiliation{Department of Physics, Institute for Advanced Studies in Basic Sciences
(IASBS),
Gava Zang,
Zanjan 45137-66731,
Iran}

\date{\today}
\begin{abstract}\label{txt:abstract}
We Have developed the concept of a new kind of single-electron transistor in which the transport of the electron through a quantum wire is controlled by charged quantum rings.  Using a 2D harmonic potential as the transverse constraint, we numerically investigated the transport of the electron through the wire.  We have shown that in the low energy limit, for a suitable configuration of the rings, called the quadrupole configuration, we are able to adjust the conductance of the wire and therefore control the switching process.
\end{abstract}

\maketitle

\section{INTRODUCTION}\label{Intro}

During the last decade there has been a growing interest for developing new electronic devices in nanoscales.  The quantum behavior of these devices gives them new abilities never seen in classical electronics.  A well-known example is the single electron transistor (SET).  It is considered  as an important element of future low power and high density integrated circuits.  SET is a nanoscale switching device which is made of an electronic channel between the source and the drain and is based on the tunneling effect.  A gate voltage $V_g$ is used to control a one-by-one electron transfer from the source to the drain via a quantum dot (QD).  When the voltage $V_g$ is less than a threshold voltage $V_t$, the system is in the Coulomb blockade state and the switch is closed.  If the voltage $V_g$ exceeds $V_t$, the switch is open and the current can start to flow through the channel.  
The goal of this paper is to introduce a new type of SET in which electronic transport in discrete channels of a quantum wire (which we call the transverse modes) is controlled by circular charged quantum rings.  This may allow better control of the switching process compared to the usual SETs.

To implement the required configuration for this study, we have to utilize quantum wires and quantum rings.  The Drude model describes the transport of electrons in conductors by the Boltzmann transport equation and introduces a mean free path $\Lambda$.  If the system's  dimensions are less than the mean free path $\Lambda$, the impurity scattering is negligible.  In this case, the electron transport can be regarded as truly ballistic (conducting without any scattering).  
At relatively low temperatures, the contribution to the conductance is given by the electrons with energies close to the Fermi surface  $E_F$.   Quantum wires are defined as electrical conductors that have a thickness or diameter $w$ comparable with the Fermi wavelength $\lambda_F$ or less, and a length $L$ less than $\Lambda$.  Therefore, the wire can be regarded as a quasi one-dimensional system in which the quantum effects influence the transport properties of the electrons.  Typical quantum wires are tens of nanometers in diameter and a few micrometers in length.  Due to their rich and fascinating electrical properties, the metallic nanowires are of particular interest \cite{Lin03}.

Let us consider a perfect quantum wire between two terminals.  We assume the quantum wire is composed of a given number of transverse modes.  According to the B\"{u}ttiker model \cite{Buttiker}, the electron dynamics in the effective mass approximation is described by the following Hamiltonian:
\begin{equation}
H=-\frac{\hbar^2}{2m^*}\nabla^2+V_c(\textbf{r}_{\perp})+E_c,
\end{equation}
where $m^*$ is the effective mass of the electron, $V(\textbf{r}_{\perp})$ is a confining potential, $\textbf{r}_{\perp}$ is in the transverse direction and $E_c$ is the conduction band edge of the (bulk) conductor material.
Because of the narrowness of the confining potential $V_c(\textbf{r}_{\perp})$ the energy for the transverse propagation is quantized ($E_{\perp}=\varepsilon_n$, $n$ is the index for the discrete spectrum).  The total energy for each transverse mode $n$ can be written as
\begin{equation}
E_n(k_z)=E_c+\varepsilon_n+\frac{\hbar^2k_z^2}{2m^*}.
\end{equation}
Here $k_z$ is the wave vector component along the wire.  

\begin{figure}
\centering
\hspace*{-5mm} 
 \includegraphics[width=1.10\columnwidth]{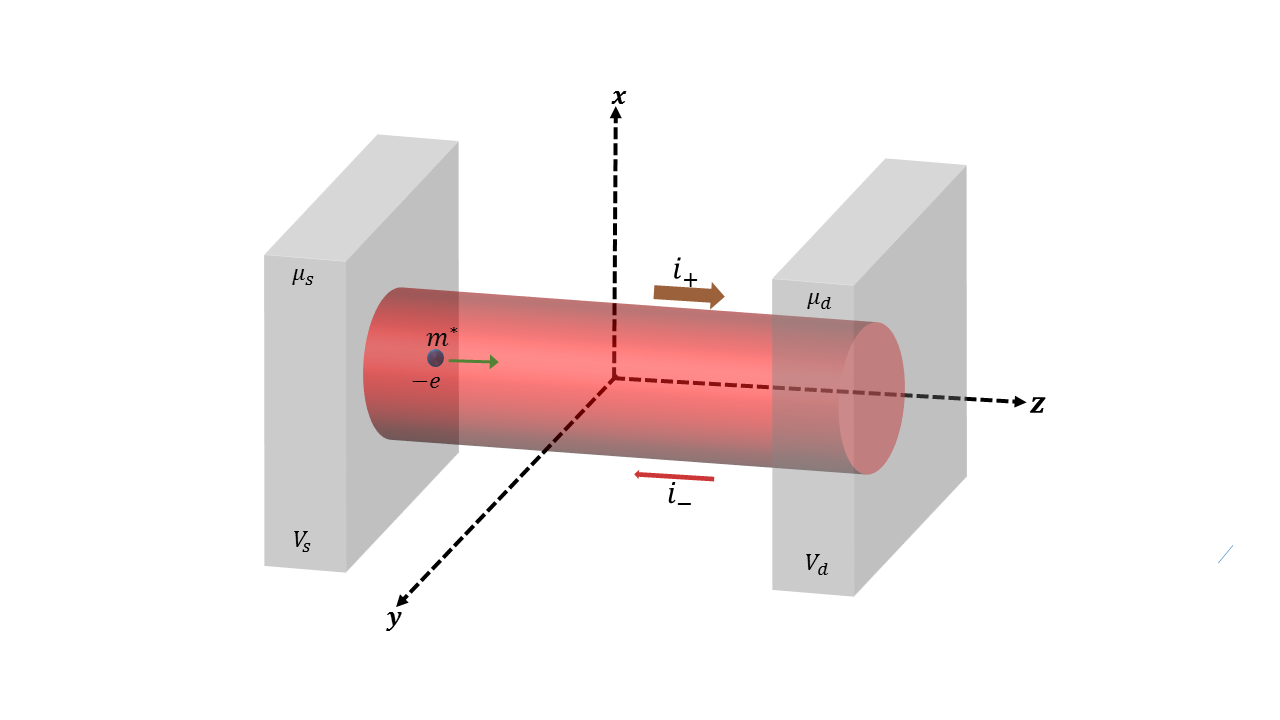}
  \label{fig1a}
  \centering
  \includegraphics[width=0.90\columnwidth]{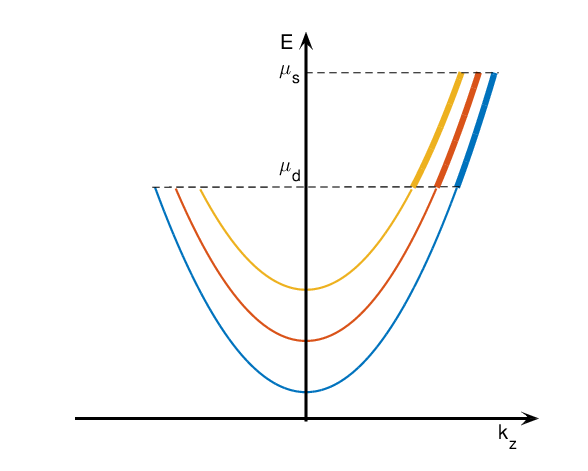}
  \label{fig1b}
\caption{Schematic drawing of energy band diagram of a perfect quantum wire connected to two contacts.}
\label{fig1}
\end{figure}

The electrons of a contact in thermodynamical equilibrium can be considered to be Fermi-distributed with an electrochemical potential $\mu$.  
Assuming a quantum wire between two contacts with chemical potentials $\mu_s$ and $\mu_d$ (Fig.\ref{fig1}), the net contribution to the conductance is from the electrons lying in the states with energy $E$ such that $\mu_d <E<\mu_s$.  In reality the potential landscape inside the device will be modified due to the electrostatic potential created by the electrons.  However, for simplicity we assume that the chemical potential is constant for $k_z>0$ and $k_z<0$ inside the channel, and we do not have to consider any voltage drop inside the device.  Besides we assume that the thermal energy $k_BT$ is much smaller than the energy gap between the levels.  In this case the conductance is given by \cite{Landauer}
\begin{equation}\label{conducotance1}
G=2R_k^{-1}M(E),
\end{equation}
where the so-called von Klitzing constant $R_k=h/e^2\approx 2.6\times 10^4\Omega$ is the standard resistance quantum and $M(E)$ is  the number of open channels of the quantum wire with an energy $E$ which is given by

\begin{equation}
M(E)=\sum_{n}\Theta(E-E_n(k_z=0)).
\end{equation}

So far we considered a perfect quantum wire i.e., without dissipation or scatterer.  However, in the case of scattering induced by a scatterer (such as defects, impurities or irregularities) the conductivity, in temperature $T$  if a small bias voltage is applied to the system ($\delta\mu=\mu_s-\mu_d\ll \mu_s$)  is given by the Landauer equation
\begin{equation}\label{conducotanceT}
G=G(E_F=\mu_s, T)=2R_k^{-1}\sum_{n, n'}\int T_{nn'}(E)\big(\frac{-\partial f}{\partial E}\big)dE,
\end{equation}
where the summation is over the open channels.  Here $T_{nn'}$ denotes the probability of transition from the state $n$ to $n'$ and $f(E,E_F)=\big[ exp(E-E_F)/k_BT+1 \big]^{-1}$ is the Fermi-Dirac distribution function.  $-\frac{\partial f(E,E_F)}{\partial ٍE}$ is referred to as the broadening function.  In zero temperature $(T=0)$ Eq.(\ref{conducotanceT}) is simplified to 
\begin{equation}\label{conducotance2}
G=2R_k^{-1}\sum_{n, n'}T_{nn'}(E_F)
\end{equation}
which in the absence of scattering ($T_{nn'}=\delta_{nn'}$) is reduced to Eq.(\ref{conducotance1}).

The other requirements for our configuration are quantum rings (QRs).  QRs and QDs are two well-known examples of quasi-zero-dimensional structures.  Because of their size which are of the order of phase-coherence length of their electrons, these structures exhibit some interesting purely quantum mechanical effects.  The electronic states of these nano-size objects are quantized just like those of an atom.  However, these so-called artificial atoms have the advantage that their electronics and optical behaviors are tunable and the engineered nature of them enables the adjusting of their properties in the fabricating process which makes them very useful.  
Zinc oxide nanorings, formed by rolling up single-crystal nano belts are well-known quantum rings \cite{Hughes}.  QRs can also be produced by droplet epitaxial growth \cite{Mano,Kuroda} and atomic force microscope tip oxidation \cite{Fuhrer, Keyser} techniques.

Due to their topology, QRs show more flexibility in electronic structure design compared to QDs.  This allows better control of their physical properties than do QDs, by varying shape parameters such as ring width and outer-inner radius ratio.
Moreover we can make coupling between electrons of the ring and a magnetic or a radiation field.  It is also possible to grow QRs on photonic crystals or optical cavities.  These abilities enable us to control behavior of the rings using quantum electrodynamical effects.   The Aharanov-Bohm effect \cite{Aharanov, Chambers} is a well-known example which arises from the direct influence of the vector potential on the phase of the electron wave function. 
This results in the magnetic-flux-like splitting of electron energy levels corresponding to mutually opposite electronic rotation\cite{Buttiker83}.
As another example one can make coupling between electrons of the ring and circularly polarized photons which results in the magnetic-flux-like splitting of the electron energy levels \cite{Kibis} and oscillations of the ring conductance as a function of the intensity and frequency of the irradiation \cite{Sigurdsson}.

QRs can be coupled to each other in different configurations, result in the formation of the so called artificial molecules.
Experimental realization \cite{Somaschini} and theoretical consideration \cite{Chwiej,Fuster,Szafran,Escartin,Planelles,Climente,Climente2006,Malet,Planelles2007,Culchac} of coupled QRs with different configurations have been reported.  The possibility to control the coupling between rings, their electron distribution, and electron transition from one ring to another have widely been considered for multiple quantum rings \cite{Kuroda,Fuhrer,Salehani,Szafran,Climente2006,Lorke2000,Keyser2002,Abbarchi,Bayer,Arsoski,Filikhin1,Filikhin2,Aronov,Chakraborty,Lee,Simonin,Voskoboynikov,Li,Cui,Sanguinetti,Chena,Szafran08,Zhu}.

\begin{figure}[tph!]
\begin{center}
\hspace*{-2cm} 
\includegraphics[width=1.30\columnwidth]{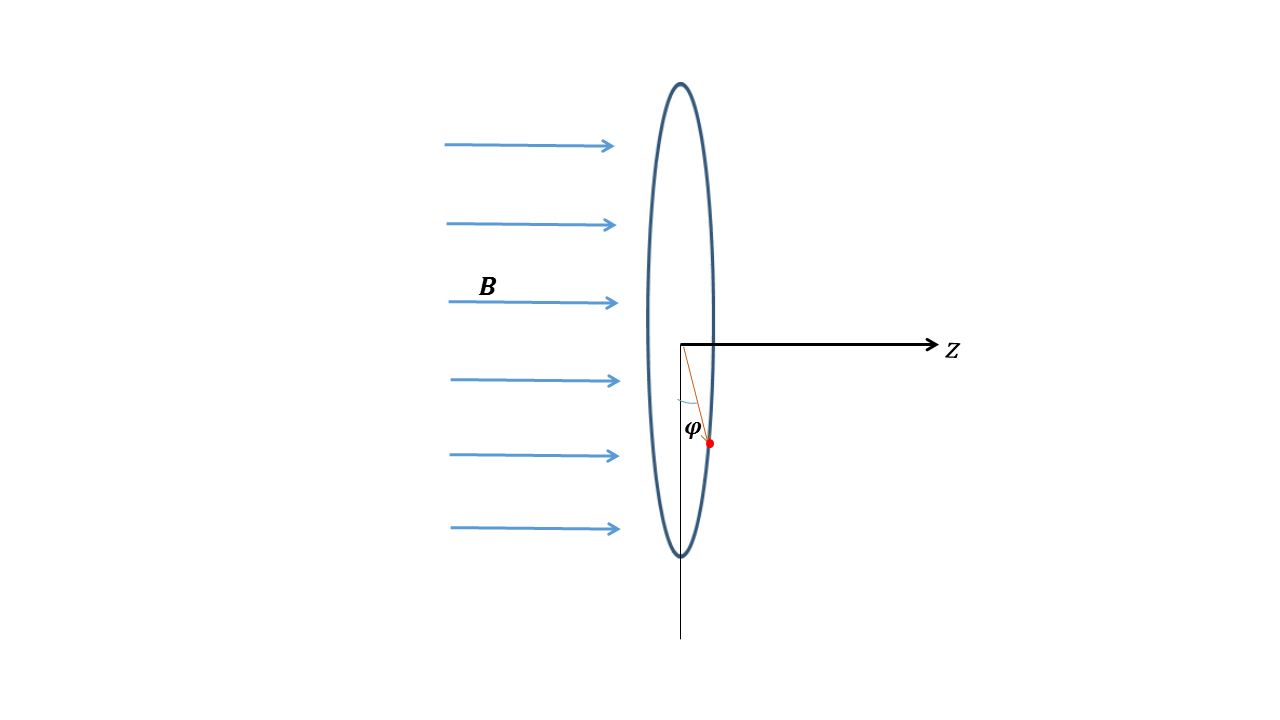}                       
\end{center}
\caption{Schematic drawing of a one-dimensional ring pierced by a magnetic flux.}
\label{fig2}
\end{figure}

\begin{figure}[tph!]
\begin{center}
\includegraphics[width=1.10\columnwidth]{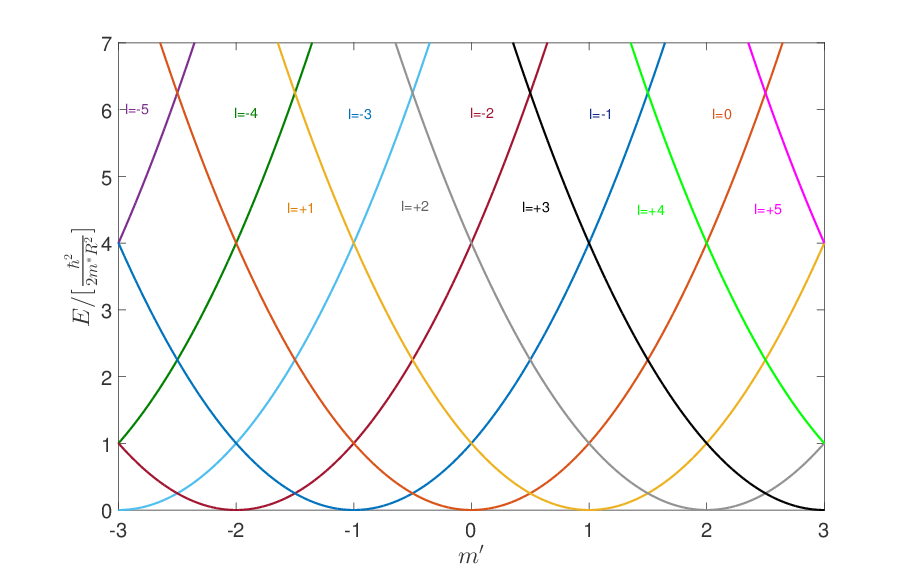}
\end{center}
\caption{Energy spectrum of the ring as a function of the number of flux quanta penetrating the ring for different values of the angular momentum $l$.} 
\label{fig3}
\end{figure}

In a simple model we can assume a ring to be perfect and strictly one dimensional with radius $R$ pierced by a magnetic flux (Fig.\ref{fig2}), and filled with spinless non-interacting electrons (only one particle per state).  In the single mode regime the 1 particle Hamiltonian in cylindrical coordinates is given by
\begin{equation}
H=\frac{\hbar^2}{2m^*R^2}\bigg(-i\frac{\partial}{\partial\phi}+m'\bigg)^2,
\end{equation}
which has  the solution 
\begin{equation}
\psi_{l, m'}(\phi)=e^{il\phi}, \quad \textrm{with}\quad E_{l, m'}=\frac{\hbar^2}{2m^*R^2}(l-m')^2,
\end{equation}
where $m'=\Phi/\Phi_0$  ($\Phi_0=h/e\approx 4.1\times 10^{-15}$ Wb is the magnetic flux quantum) is the number of flux quanta penetrating the ring and $l$ is the angular momentum of the electron which is integer due to the boundary condition $\psi_{l,m'}(\phi+2\pi)=\psi_{l,m'}(\phi)$.  The spectrum of this Hamiltonian has been shown in Fig.\ref{fig3}.  It consists of parabolas shifted with respect to each other by one flux quanta.  Each parabola corresponds to a certain value of $l$. 


For $N$ electrons the (trivial) total energy and total angular momentum are 
\begin{equation}
E=\sum_{i}^{N}E_{l_i,m'}=\sum_{i}^{N}\frac{\hbar^2}{2m^*R^2}(l_i-m')^2,
\end{equation}
and 
\begin{equation}
L_z=\sum_{i}^{N}l_i\hbar,
\end{equation}
respectively.  Here we assumed to be no Coulomb interaction between the electrons.  However this interaction can be described by the capacitance $C$ of the system.  Actually given a certain number $N$ of electrons the energy that it takes to overcome the Coulomb repulsion in order to bring one more electron is given by
\begin{equation}
E_{add}=\frac{Ne^2}{C}+\Delta E,
\end{equation}
where $\Delta E$ is the difference of the quantum levels of the system.  The number of electrons on the ring can be tuned one by one (which is a practical requirement for our configuration of SET), e.g. by adjusting the voltages of the ring \cite{Ihn05,Ihn03,Xiang}.  Due to the Coulomb blockade, just at certain voltages (corresponding to constructive interference of the electron wave function, which results in high conductance of the ring), the provided energy is sufficient to compensate energy difference of the $N$ and $N+1$ many particle states in the ring and induce another electron on the ring \cite{Ihn03}.

This paper is organized as follows: Sec.\ref{Intro} provides an introduction to the subject.  Our model for SET is described in Sec.\ref{Model}.  In Sec.\ref{Numeric} we explain the numerical approach.  Our results are shown in Sec.\ref{Results}.  Finally we summarize and conclude with Sec.\ref{Summary}.

\section{Our Model for SET}\label{Model}
\begin{figure}[tph!]
\begin{center}
\hspace*{-15mm}
\includegraphics[width=1.30\columnwidth]{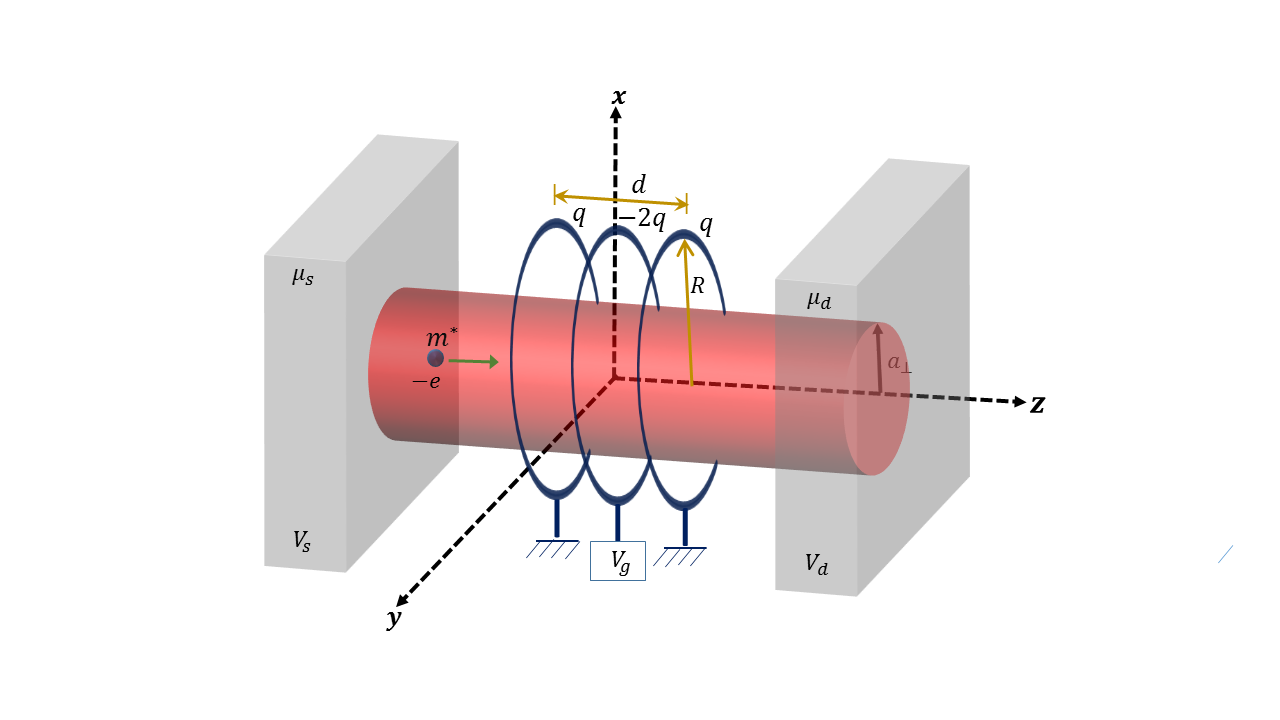}
\end{center}
\caption{Schematic drawing of our configuration for SET.}
\label{fig4}
\end{figure}
To expand our work more clearly, let us consider the schematic drawing of the configuration as shown in Fig.\ref{fig4}: a moving electron passing through a quantum wire surrounded by a triple quantum ring (TQR).  The TQR consists of three coaxial quantum rings acting as an electrostatic quadrupole and applying an external interaction potential on the electron.  Assuming the charge distribution of the rings (q for the side rings and -2q for the inner ring) to be homogeneous for simplicity, some straightforward calculations for the interaction potential leads to \cite{Jackson}
\begin{multline}\label{elecpot}
V(\boldsymbol{r})=\frac{-2eq}{4\pi\varepsilon_0}\sum_{l=0}^{\infty}P_{2l}(\cos\theta)\Bigg[\frac{r_{\textless}^{2l}}{r_{\textgreater}^{2l+1}}P_{2l}(\cos\alpha)\\
-\frac{r'^{2l}_{\textless}}{r'^{2l+1}_{\textgreater}}P_{2l}(0)\Bigg],
\end{multline}
where $r_{\textless}(r_{\textgreater})=\min (\max)\left\{r, [R^2+(d/2)^2]^{1/2}\right\}$, $r'_{\textless} (r'_{\textgreater})=\min(\max)\left\{r, R\right\}$, $\cos\alpha=d/[4R^2+d^2]^{1/2}$ and $P_l(x)$s are the Legendre polynomials.
$R$ and $d$ are the radius and length of the TQR, respectively.  We chose the 2D harmonic oscillator with frequency $\omega$ as the transverse constraint for the quantum wire.
The related radius of  the wire can be approximated as $a_{\perp}=(\hbar/m^*\omega)^{1/2}$.   $m^*$ is the effective mass of the electron.  The electron moving through the wire feels two originally different potentials: the external interaction potential owing to the charged rings and the 2D harmonic potential of the confinement.  

In the effective mass approximation the Schr\"{o}dinger equation governing the motion of the electron with mass $m^*$ and energy $E=E_c+E_{\perp}+E_{||}\approx E_F$, is
\begin{equation}
\left[-\frac{\hbar^2}{2m^*}\triangledown^2+V(\boldsymbol{r})+\frac{1}{2}m^*\rho^2\omega^2+E_c\right]\psi(\boldsymbol{r})=E\psi(\boldsymbol{r}).
\end{equation}
Here $E_{||}\approx E_F$ and $E_{\perp}=\hbar\omega(2n+|m|+1)$ are the longitudinal and transverse collision energies respectively.  The radial and azimuthal quantum numbers $n$ and $m$ of the 2D harmonic oscillator independently takes the values $n=0, 1, 2, ...$ and $m=0,\pm 1, \pm 2,...$, respectively. Performing the scale transformation $r\rightarrow\frac{r}{a_0}$, $E\rightarrow\frac{E}{E_0}$, $\omega\rightarrow\frac{\omega}{\omega_0}$ and $V(\boldsymbol{r})\rightarrow\frac{V(\boldsymbol{r})}{E_0}$ (here $a_0$ is an arbitrary constant, $E_0=\hbar^2/(m^*a_0^2)$ and $\omega_0=E_0/\hbar$; we assume that $a_0=2$nm and $m^*=0.067 m_e$) leads to the scaled dimensionless Schr\"{o}dinger equation
\begin{equation}\label{SchEq}
\left[-\frac{1}{2}\triangledown^2+V(\boldsymbol{r})+\frac{1}{2}\rho^2\omega^2+E_c\right]\psi(\boldsymbol{r})=E\psi(\boldsymbol{r}).
\end{equation}
Due to the axial symmetry of the system, the angular momentum component along the z-axis ($m\hbar$) is conserved and one can separate the $\phi-$variable. Therefore the problem can be reduced to a 2D one.  In the asymptotic region, $|z|\rightarrow\infty$, the electric potential (\ref{elecpot}) falls off as $r^{-3}$ and thus it is negligible compared to other terms of the Hamiltonian.  In this case the axial and transverse motions decouple (this is the reason why we use quadrupole instead of monopole or dipole configuration).  Thus the quantum numbers $n$ and $m$ can be used to label the asymptotic states (channels) as $e^{ik_nz}\Phi_{n, m}(\rho, \phi)$, where $\Phi_{n,m}(\rho,\phi)$s are the eigenfunctions of the 2D harmonic potential, and the momentum $k_n$ is defined as 
\begin{equation}
k_{n}=2\sqrt{\frac{E'}{2} - n - \frac{|m|}{2}},
\end{equation}
with $E'=E-E_c-\omega$.

Assuming the electron to be initially in the channel $n$, the asymptotic wave function (when $z\to \pm\infty$) takes the form 
\begin{equation}\label{asymp}
\psi_{n,m}(\boldsymbol{r})=
e^{ik_nz}\Phi_{n,m}(\rho,\phi)+\sum_{n'=0}^{n_e}f^{\pm}_{nn'}e^{ik_{n'}|z|}\Phi_{n',m}(\rho,\phi),
\end{equation}
where $f^{\pm}_{nn'}$s are the scattering amplitudes which describe transitions between the channels $n$ and $n'$. $n_e$ is the number of the open excited transverse channels, i.e. the maximum value of $n'$ such that  $k_{n'}$ is real.
The scattering amplitude depends also on $m$.  However, $m$ is conserved.  Hereafter we assume $m=0$.

Using the asymptotic wave function (\ref{asymp}) one obtains for the inelastic transmission coefficient $T_{nn'}$ 
\begin{equation}\label{Trans}
T_{nn'}=\Theta(\epsilon-n')\frac{k_{n'}}{k_n}|\delta_{n,n'}+f^+_{nn'}|^2,
\end{equation}
where $\Theta(x)$ is the Heavyside step-function.

\section{Numerical Approach}\label{Numeric}

To solve Eq.(\ref{SchEq}) we use the method introduced in \cite{Saeidian}.  First we discretize (\ref{SchEq}) on a grid of the angular variable $\{\theta_j\}^{N_{\theta}}_{j=1}$  with $\theta_j$ being the roots of $P_{N_{\theta}}(\cos\theta)$ and expand the solution $\psi(r, \theta)$ as
\begin{equation}\label{expwave}
\psi(r,\theta)=\sum_{j=1}^{N_{\theta}}\frac{u_j(r)}{r}g_j(\theta),
\end{equation}
where $g_j(\theta)=\sum_{l=0}^{N_{\theta}-1}P_l(\cos\theta)A_{lj}$.  In order for $\psi(r, \theta)$ to be finite at the origin, we must have 
\begin{equation}\label{BouConOrigin}
u_j(r=0,\theta)=0.
\end{equation}
The coefficients $A_{lj}$ are defined as $A_{lj}=[\hat{P}^{-1}]_{lj}$ and $[\hat{P}^{-1}]$ is the inverse of the $N_{\theta}\times N_{\theta}$ matrix $[\hat{P}]$ with elements $[\hat{P}]_{lj}=\lambda_jP_l(\cos\theta_j)$.  $P_l(x)$s are the normalized Legendre Polynomials and $\lambda_j$s are the weights of the Gauss quadrature.  By substituting (\ref{expwave}) into (\ref{SchEq}) we arrive at a system of $N_{\theta}$ coupled equations
\begin{multline}
\big[-\frac{d^2}{dr^2}+2(V(r,\theta_j)+\frac{1}{2}\omega^2 r^2\sin^2\theta_j-E)\big]u_j(r)+\\
\frac{1}{r^2}\sum_{j'=1}^{N_{\theta}}\sum_{l=0}^{N_{\theta}-1}\lambda_j^{\frac{1}{2}}\lambda_{j'}^{\frac{1}{2}}l(l+1)P_l(\cos\theta_j)A_{lj'}u_{j'}(r)=0
\end{multline}

By mapping and discretizing $r\in(0, r_m]$ onto the uniform grid $x_j\in(0,1]$ according to $r_j=r_m\frac{e^{\gamma x_j}-1}{e^{\gamma}-1}$, with $j=1, 2, ..., N$ (here $r_m$ is chosen in the asymptotic region $r\rightarrow\infty$, and $\gamma$ is a tuning parameter) and using the finite difference approximation we solve the above system of equations with the boundary conditions (\ref{asymp}) in a modified form (in which we eliminate the unknown coefficients $f^{\pm}$) and (\ref{BouConOrigin}), for fixed colliding energy $E$.  By matching the calculated wave function $\psi(r, \theta)$ with the asymptotic behavior (\ref{asymp}) at $r=r_m$, we find the scattering amplitudes $f^{\pm}_{nn'}$ from which we can calculate $T_{nn'}$ (see Eq.(\ref{Trans})). 

\section{Results}\label{Results}

 \begin{figure}[tph!]
\begin{center}
\includegraphics[width=0.90\columnwidth]{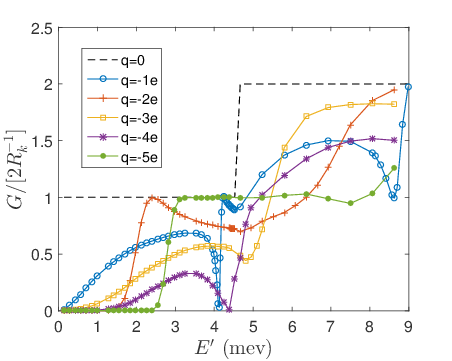}
\end{center}
\caption{The conductance $G$ of the quantum wire as a function of $E'$ for upto two open channel for different values of $q$ for $R=30$nm, $d=10$nm and $a_{\perp}=20$nm.}
\label{fig5}
\end{figure}

\begin{figure}[tph!]
\begin{center}
\includegraphics[width=0.90\columnwidth]{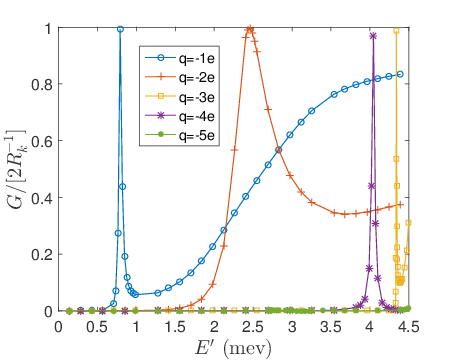}
\end{center}
\caption{The plot of $G$ vs. $E'$ in one-mode regime for $R=30$nm, $d=20$nm and $a_{\perp}=20$nm and for different values of $q$.}
\label{fig6}
\end{figure}

\begin{figure}[tph!]
\begin{center}
\includegraphics[width=0.90\columnwidth]{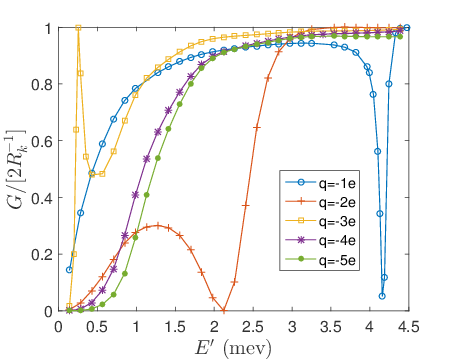}
\end{center}
\caption{The conductance $G$ of the quantum wire as a function of $E'$ for different values of $q$ for $R=40$nm, $d=10$nm and $a_{\perp}=20$nm in one-mode regime.}
\label{fig7}
\end{figure}

In Fig.\ref{fig5} we have depicted $G$ vs $E'$ for up to two open channels for different values of $q$.  The results have been obtained for $R=30$nm, $d=10$nm and $a_{\perp}=20$nm.  It is obvious that $G$ increases as the number of open channels increases.  However, there are some minimums which are due to the so-called confinement induced resonance (CIR).  CIR was first studied for ultracold atoms in harmonic waveguides by M. Olshanii in 1998 \cite{Olshanii}.  He showed that CIR for s-wave scattering is a kind of the zero-energy Feshbach resonance.  It occurs when the total energy of the incident particle in the confined geometry coincides with the energy of the quasi-bound state of the next excited channel.  In this case the effective force experienced by the particle diverges and the transmission coefficient $T$, reaches a minimum, which is zero in single mode regime (i.e. the particle is totally reflected).   CIR has been studied for higher partial waves as well \cite{Granger,Kim,Saeidian, Giannakeas}.  We observe the same effect in our problem: the confining potential induces a resonance which is responsible for minimization of the transmission coefficient $T$. 

We see from Fig.\ref{fig5} that for $q=-2e$  as well as $q=-5e$, $G$ remains almost zero (no current) in a relatively wide range of $E'$ (i.e., the switch is closed).  With increasing $E'$, the conductance $G$ increases rather abruptly by 1 unit.  

On the other hand, for fixed $E'$ we can control the switching process by tuning the applied voltages of the rings (i.e., changing the charges of the rings).

Fig.\ref{fig6} shows a graph of $G$ vs. $E'$ in the one-mode regime.  All the parameters in Fig.\ref{fig6} are the same as those in Fig.\ref{fig5} except for $d$ ($=20$nm) which is larger.  There is a peak for each value of $q$.  However, $G$ is smaller on average compared with Fig.\ref{fig5}.  For $q=-3e$ and $q=-4e$, $G$ is almost zero except for a sharp peak at large $E'$.  This may allow us to measure small potential difference $\delta V=V_s-V_d$ between the source and drain.

\begin{figure}[tph!]
\begin{center}
\includegraphics[width=0.90\columnwidth]{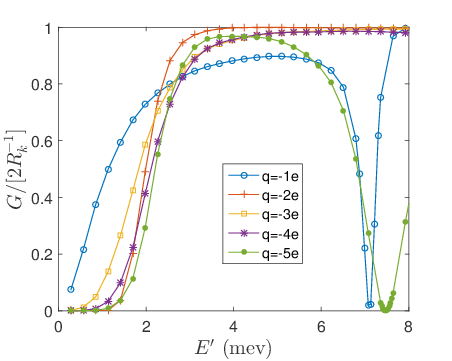}
\end{center}
\caption{The conductance $G$ of the quantum wire as a function of $E'$ in one-mode regime for $R=30$nm, $d=10$nm and $a_{\perp}=15$nm for different values of $q$ .}
\label{fig8}
\end{figure}

Figs.\ref{fig7} and \ref{fig8} shows our results in single mode regime for the same parameters as those in Fig.\ref{fig5} but for larger $R$ ($=40$nm) and $a_{\perp}$ ($=15$nm), respectively.  The conductance $G$ tends to increase when we increase the energy of the electron, remaining nearly constant in its maximum in a relatively wide range of $E'$.  However it falls into zero rather abruptly in certain energies.  Comparing these figures with Fig.\ref{fig5} shows dependence of the conductance on $R$ and $a_{\perp}$.  The larger $R$ or the smaller $a_{\perp}$, the larger on average the conductance is.

\begin{figure}[tph!]
\begin{center}
\includegraphics[width=0.90\columnwidth]{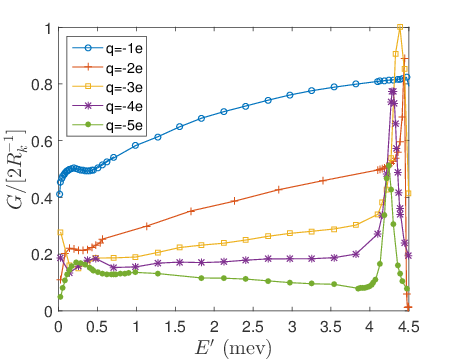}
\end{center}
\caption{The plot of $G$ as a function of $E'$ in one-mode regime for $R=30$nm, $d=10$nm and $a_{\perp}=20$nm for different values of $q$.  Here $q$ is positive.}
\label{fig9}
\end{figure}

\begin{figure}[tph!]
\begin{center}
\includegraphics[width=0.90\columnwidth]{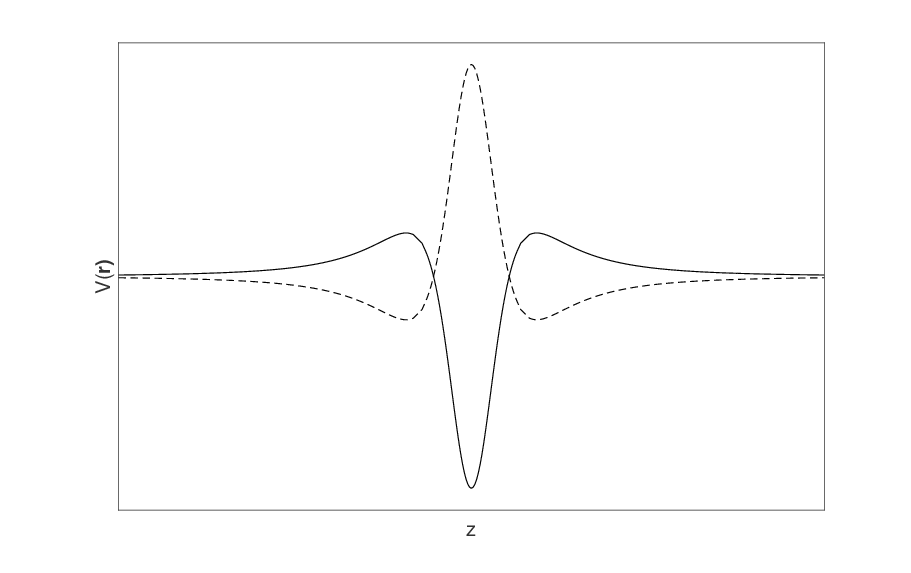}
\end{center}
\caption{Schematic graph of the interaction potential $V(\boldsymbol{r})$ along the $z-$axis (x=y=0) for $q< 0$ (solid line) and $q > 0$ (dashed line).}
\label{fig10}
\end{figure}

So far we have considered the case $q< 0$, i.e., the side rings have gained electrons while the inner ring has lost electrons.  Fig.\ref{fig9} shows $G$ vs $E'$ for the reverse case ($q > 0$) for the same parameters as Fig.\ref{fig5}.  In this case $G$ changes very slowly as $E'$ increases except for a sharp peak at large $E'$ near to the next channel threshold.  The peak shifts to the left as $q$ increases.  The larger $q$, the smaller on average the conductance is.  This is because of the fact that for larger $q$ the moving electron experiences a stronger potential barrier  (see Fig.\ref{fig10}).  The stronger the potential barrier, the less the probability of tunneling is. 

To explain the dependence of $G$ on the sign of $q$ we have plotted a schematic diagram of the interaction potential $V(\bf{r})$  along the $z$-axis in Fig.\ref{fig10}.  The shape of the potential depends on the sign of $q$.  For $q<0$ there is a strong well between two weak barriers.   The potential may support quasi-bound state leading to CIR.  While for $q>0$ we see two shallow minimums separated by a strong potential barrier.   The electron transmission is based on tunneling through this barrier.

\section{SUMMARY AND CONCLUSION}\label{Summary}
In summary, we have introduced a new kind of single electron transistor based on quadrupole configuration of quantum rings.     With setting the charges of the rings, we can control conductance of the wire.  Due to the Coulomb blockade, the charges of the rings can be changes just in specific voltages which can be engineered by applying magnetic field and/or circularly polarized light.  This allows better control of the switching process compared to the usual single electron transistor based on quantum dots.

\section{ACKNOWLEDGMENTS}
Sh. S would like to thank A. Valizadeh and A. Ghorbanzadeh for fruitful discussions and S. Azizi for his contribution to the figures.

\end{document}